\documentclass[prd,amsfonts,nofootinbib,showpacs]{revtex4}
\usepackage{graphicx}
\usepackage{dcolumn}
\usepackage{bm}
\usepackage{amsmath}
\usepackage{epsfig}
\usepackage{color} 
\usepackage{multirow} 

\renewcommand{\baselinestretch}{1.25}

\begin{document}

\title{ Low-energy neutrino-electron scattering as a Standard Model
  probe :\\ the potential of LENA as case study}

\author{E. A. Garc\'es}%
\email{egarces@fis.cinvestav.mx} 
\author{O. G. Miranda}%
\email{Omar.Miranda@fis.cinvestav.mx}
\affiliation{Departamento de F\'\i sica, Centro de Investigaci\'on 
  y de Estudios Avanzados del IPN,\\
  Apdo. Postal 17-740, 07000, Mexico DF, MEXICO}

\author{M. A. T\'ortola}%
\email{mariam@ific.uv.es}
\author{J. W. F. Valle}
\email{valle@ific.uv.es}
\affiliation{ AHEP Group, Institut de F\'{\i}sica Corpuscular --
 C.S.I.C./Universitat de Val{\`e}ncia \\
 Edificio Institutos de Paterna, Apt. 22085, E--46071 Valencia, Spain }

\begin{abstract}

  Several proposals for studying neutrinos with large detectors are
  currently under discussion. We suggest that they could provide a
  precise measurement of the electroweak mixing angle as well as a
  probe for new physics, such as non-standard neutrino interactions
  (NSI), and the electroweak gauge structure.  We illustrate this
  explicitly for the case of the LENA proposal, either with an
  artificial radioactive source or by using the solar neutrino flux.

\end{abstract}
\pacs{13.15.+g,  12.15.-y, 12.60.Cn, 14.60.St }
\maketitle

\section{Introduction}

The historic discovery of neutrino
oscillations~\cite{nakamura2010review} implies that neutrinos are
massive and, therefore, the Standard Model of elementary particles
should be extended~\cite{Schwetz:2011zk}. The nature of the required
new physics remains elusive but there are strong experimental and
theoretical efforts to shed light on the correct roadmap.
 
A new generation of proposed large neutrino experiments involving
different techniques such as liquid scintillators, water Cerenkov and
liquid Argon detectors is currently under the R\&D
phase~\cite{Rubbia:2010zz}. 
Experiments such as LENA~\cite{Wurm:2011zn},
DAEdALUS~\cite{Alonso:2010fs} or MEMPHYS~\cite{deBellefon:2006vq}
could serve as multi-purpose experiments to improve our current
knowledge of neutrino oscillation parameters as well as to test
physics beyond the Standard Model (SM).

Low energy neutrino experiments provide a clean way to probe the weak
mixing angle, for example in reactor neutrino
experiments~\cite{Conrad:2004gw} or in arrays of water Cerenkov
detectors~\cite{Agarwalla:2010ty}, with an expected sensitivity in the
range of few percent or less.
While precise determinations of $\sin^2\theta_{\rm W}$ in the high
energy regime exist, the situation changes when going to lower
energies. Even for the case of neutrino nucleon scattering the NuTeV
collaboration~\cite{Zeller:2001hh} reported a discrepancy of the
expected value for $\sin^2\theta_{\rm W}$.  Recent studies suggest the
need for a re-estimation of the systematical
errors~\cite{Bentz:2009yy,Ball:2009mk}, leading to an error for
$\sin^2\theta_{\rm W}$ of the order of
$1-5$\%~\cite{Bentz:2009yy,Ball:2009mk}.  For the case of neutrino
and anti-neutrino scattering off electrons this situation is worse and
the current accuracy in the determination of the weak mixing angle is
about $10-20$\%~\cite{Deniz:2009mu,Auerbach:2001wg,Barranco:2007ea}.

It is also of great interest to investigate the potential sensitivity
of low-energy neutrino electron scattering experiments to various
types of new physics, such as non-standard interactions potentially
associated to the mechanism of neutrino mass
generation~\cite{schechter:1980gr,Maltoni:2004ei} and/or new gauge
bosons~\cite{mohapatra:1980yp,valle:1987sq,malinsky:2005bi}. Indeed
there have been suggestions in this direction~\cite{Miranda:1997vs},
as well as proposals to test possible oscillations of active neutrinos
into sterile ones~\cite{Vergados:2011na}.

Here we study the potential of the LENA proposal~\cite{Wurm:2011zn}
towards an improved measurement of the the weak mixing angle. We focus
on the case of an artificial radioactive neutrino source as has been
considered in the proposal. We find that this experimental setup could
bring an improvement in the sensitivity to $\sin^2\theta_{\rm W}$ in
this range of energies.  We also discuss how this setup could probe
physics beyond the Standard Model. In case the LENA proposal will
operate without the artificial neutrino source the potential
sensitivity on the measurement of $\sin^2\theta_{\rm W}$ is reduced,
although we speculate on the possible use of the Beryllium solar
neutrino signal to make this measurement.

\section{LENA potential  with an artificial neutrino
  source}
\label{sec:lena-potential-with}

Since the proposed LENA detector is 100 m long, one may, at least in
principle, detect different neutrino rates at different distances from
the source inside the detector~\cite{Wurm:2011zn}. Indeed, the LENA
proposal has considered the possibility of using an artificial
radioactive neutrino source to perform neutrino oscillation
measurements at short baselines, especially oscillometry tests for
sterile neutrino conversions.
Here we focus on an alternative application of such a source, namely,
the precise determination of the neutrino electron cross section and
therefore (A) the possible determination of the electroweak mixing
angle and also (B) the sensitivity to new physics such as NSI and/or
additional neutral gauge bosons.

\subsection{Sensitivity to the electroweak mixing angle}
\label{sec:sens-electr-mixing}

Although the weak mixing angle has been measured with extraordinary
precision, this is not the case for leptonic processes, especially for
the case of low energy experiments. To cite an example, a recent
determination of this SM parameter from anti-neutrino electron
scattering off electron reported the value~\cite{Deniz:2009mu}
$\sin^2\theta_{\rm W} = 0.251 \pm 0.031{\rm (stat)}\pm 0.024{\rm
  (sys)}$.

Within the Standard Model the $\nu_e e$ differential cross section is
given by
\begin{equation}
\frac{d\sigma}{dT} = \frac{2 G_F m_e}{\pi}
\big[ 
g^2_L + g^2_R(1 - \frac{T}{E_\nu})^2 - g_L g_R \frac{m_e T}{E^2_\nu},
\big]
\label{diff:cross:sec}
\end{equation}
where $G_F$ is the Fermi constant, $m_e$ is the electron mass, $T$ is
the kinetic energy of the recoil electron and $E_\nu$ is the neutrino
energy.  The coupling constants $g_L$ and $g_R$ at tree level can be
expressed as
\begin{eqnarray}
\label{gLgR}
g_L &=& \frac12 + \sin^2\theta_W\\
g_R &=&    \sin^2\theta_W .
\end{eqnarray}
Radiative corrections to the $\nu_e e$ process could give a correction
to these coupling constants of about $2$\%~\cite{Bahcall:1995mm}.
Throughout this paper we will write the tree level expressions in order
to make the discussion more transparent, however, the radiative
corrections will always be included in our computations following the
expressions discussed in~\cite{Bahcall:1995mm} with the more recent
estimate for the weak mixing angle value $\sin^2\theta_W=
0.2313$~\cite{nakamura2010review}.

In order to obtain a better determination for $\sin^2\theta_{\rm W}$
we consider here the particular setup of a $^{51}$Cr source of
$5$~MCi intensity with a monochromatic neutrino line at $747$~keV,
considered in the LENA proposal~\cite{Wurm:2011zn}.  During the
half-life of the source (28 days), the neutrino flux would give a
signal of about $1.9\times10^5$ neutrinos.

We consider two different scenarios. In the first case we estimate the
sensitivity by assuming the total number of events of the detector in
the full recoil electron energy range from $200$ to $550$~keV , while
in the second case we study the possibility of an analysis in seven
bins of $50$~keV width. That is, we take the neutrino events to be
given by
\begin{equation}
N_{i} = n_e\phi_{Cr}\Delta t \int^{T_{i+1}}_{T_i}\int  
                    \frac{d\sigma}{dT}  R(T,T') dT' dT,
\label{diff:cross:sec-sm}
\end{equation}
where $n_e$ is the number of electron targets, $\phi_{Cr}$ is the
neutrino flux coming from the $5$~MCi neutrino source, $\Delta t$ is
the $28$ days time window which corresponds to the half-life of the
source, and the resolution function $R(T,T')$ accounts for the
distribution of the measured recoil electron energy, $T$, around the
true energy $T'$:
\begin{equation}
R(T,T') = \frac{1}{\sigma\sqrt{2\pi}}\exp[-\frac{(T-T')^2}{(2\sigma^2)}],
\end{equation}
where $\sigma = 0.075\sqrt{T/MeV}$ is the expected energy resolution. 

\begin{figure}
\includegraphics[width = 0.45 \textwidth]{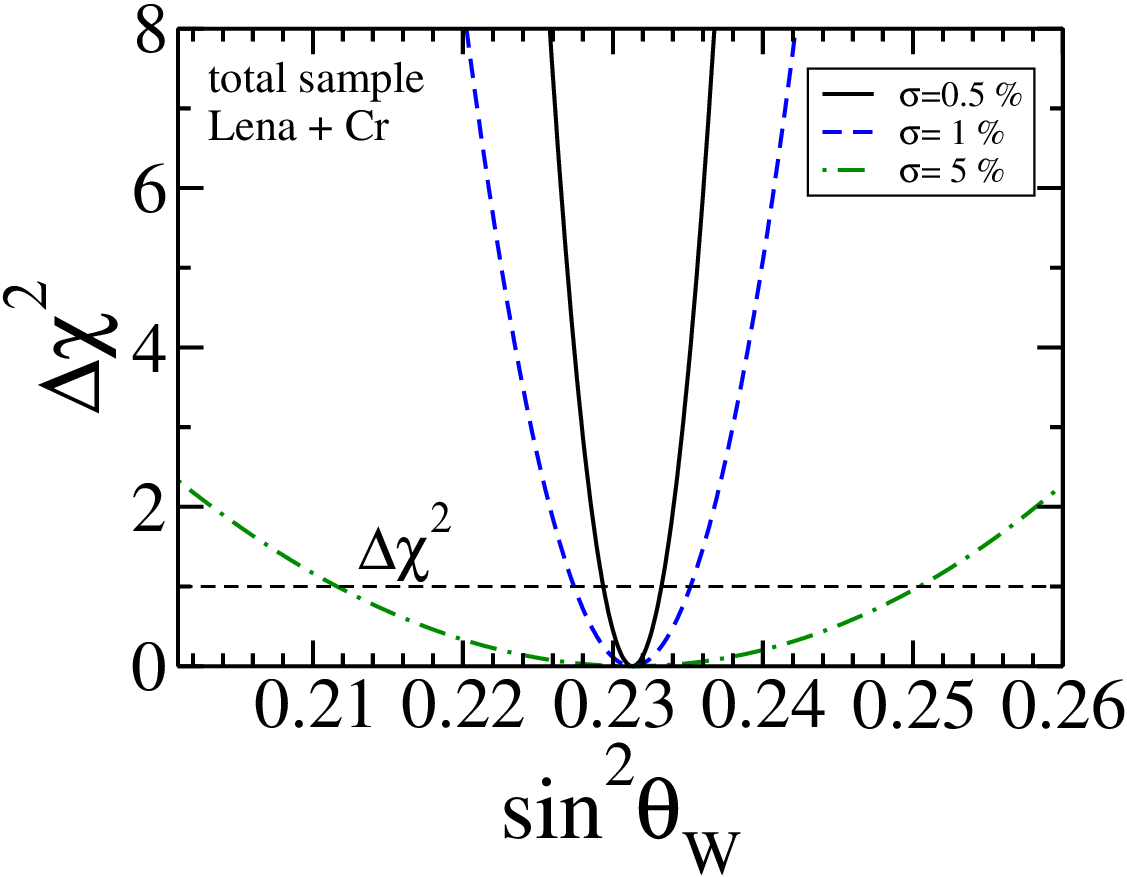}
\includegraphics[width = 0.45 \textwidth]{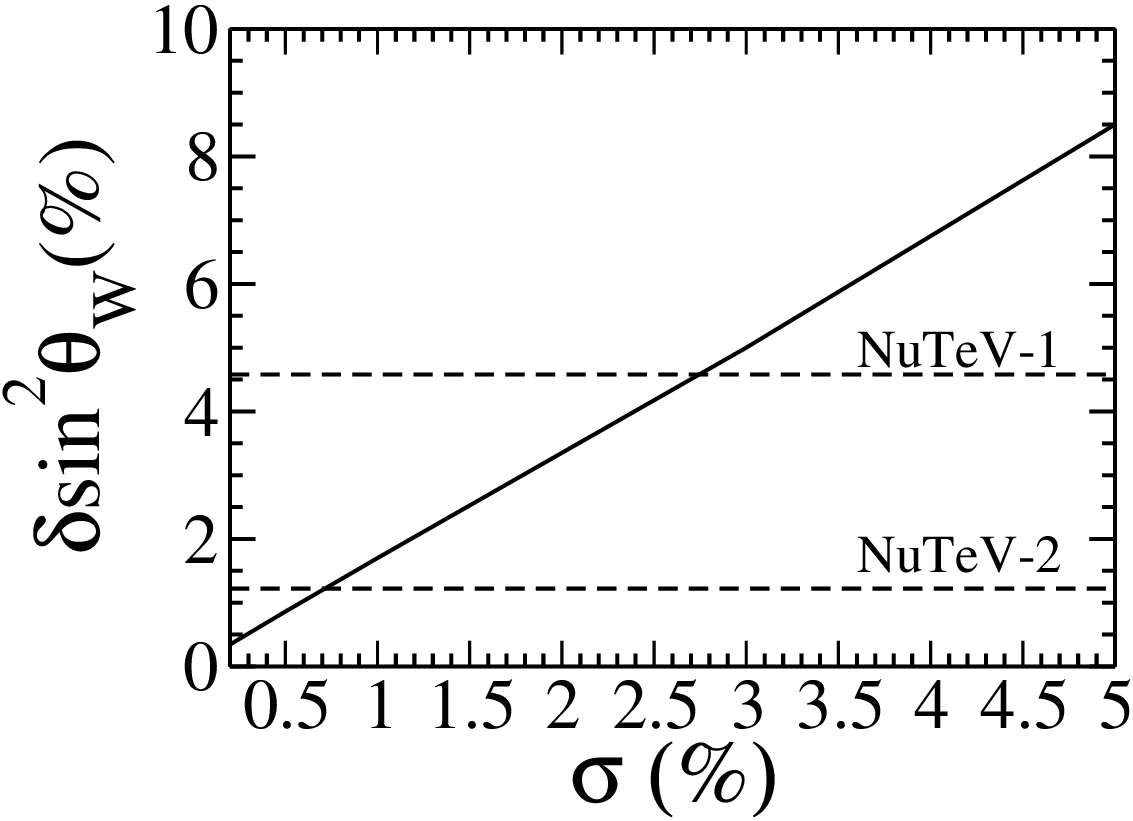}
\caption{Expected sensitivity to the electroweak mixing angle of a
  $^{51}$Cr neutrino source with the LENA detector.  We show the
  result of a simulated $\chi^2$ analysis of the total number of
  events with a given total 'error' of $0.5$, $1$ and $5$\% (left
  panel). We also show the expected precision at 1$\sigma$ (68.27
  \% CL) on $\sin^2\theta_{\rm W}$ as a function of a given percentual
  error (tilted line in right panel). Current NuTeV sensitivities
  associated to two evaluations of the systematical
  errors~\cite{Bentz:2009yy,Ball:2009mk} are also shown as horizontal
  lines.}
\label{fig:total-Cr}
\end{figure}

As already mentioned, we take two different scenarios. In the first
one we consider the whole recoil electron energy window from $250$ to
$550$~keV and therefore we have only one bin that collects all the
$1.9\times10^5$ expected events, with a small statistical error,
around $0.2$\%. In practice, one expects a larger error due to
systematics and therefore, although we can not forecast the future
precision of the experiment, we can perform our computation for
different errors and determine the corresponding precision in the
measurement of $\sin^2\theta_{\rm W}$.

To perform these computations we first assume that the detector will
measure exactly the SM prediction and perform an ideal $\chi^2$
analysis assuming a given error for the data. With this input we have
performed a $\chi^2$ analysis that gives us an idea of the sensitivity
of LENA to a new measurement of the weak mixing angle with the help of
the function
\begin{equation}
\chi^2 = \sum_{i} \frac{(N^{\rm theo}_i - N^{\rm exp}_i)^2}{\sigma^2_i},
\end{equation}
where $N^{\rm theo}_i$ is the expected number of events for different
values of $\sin^2\theta_{\rm W}$ for a given bin $i$, $N^{\rm exp}_i$
is the 'experimental' value given by the expected number of events for
the SM prediction and $\sigma^2_i$ is the error per bin (which in this
first scenario corresponds to one total bin). In order to estimate the
future LENA sensitivity we have assigned different values to the error
$\sigma^2_i$.

We show the results of our analysis in Fig.~\ref{fig:total-Cr}. In
the left panel of this figure we can see the corresponding $\chi^2$
function for three different values of the experimental error ($0.5$,
$1$ and $5$ percent).
The tilted line in the right panel indicates the expected sensitivity
to the weak mixing angle at 1$\sigma$ as a function of the assumed
experimental error. In the same right panel we show the current NuTeV
sensitivities corresponding to two evaluations of the systematical
errors of the experiment~\cite{Bentz:2009yy,Ball:2009mk}.
From the right panel one sees that an experimental error of the order
of 2.5\% would be required in order to improve the sensitivity
obtained by the NuTeV-1 result, while only an experimental uncertainty
smaller than 0.7\% would improve the results given by the second NuTeV
determination of the electroweak mixing angle.
\begin{figure}
\includegraphics[width = 0.45 \textwidth]{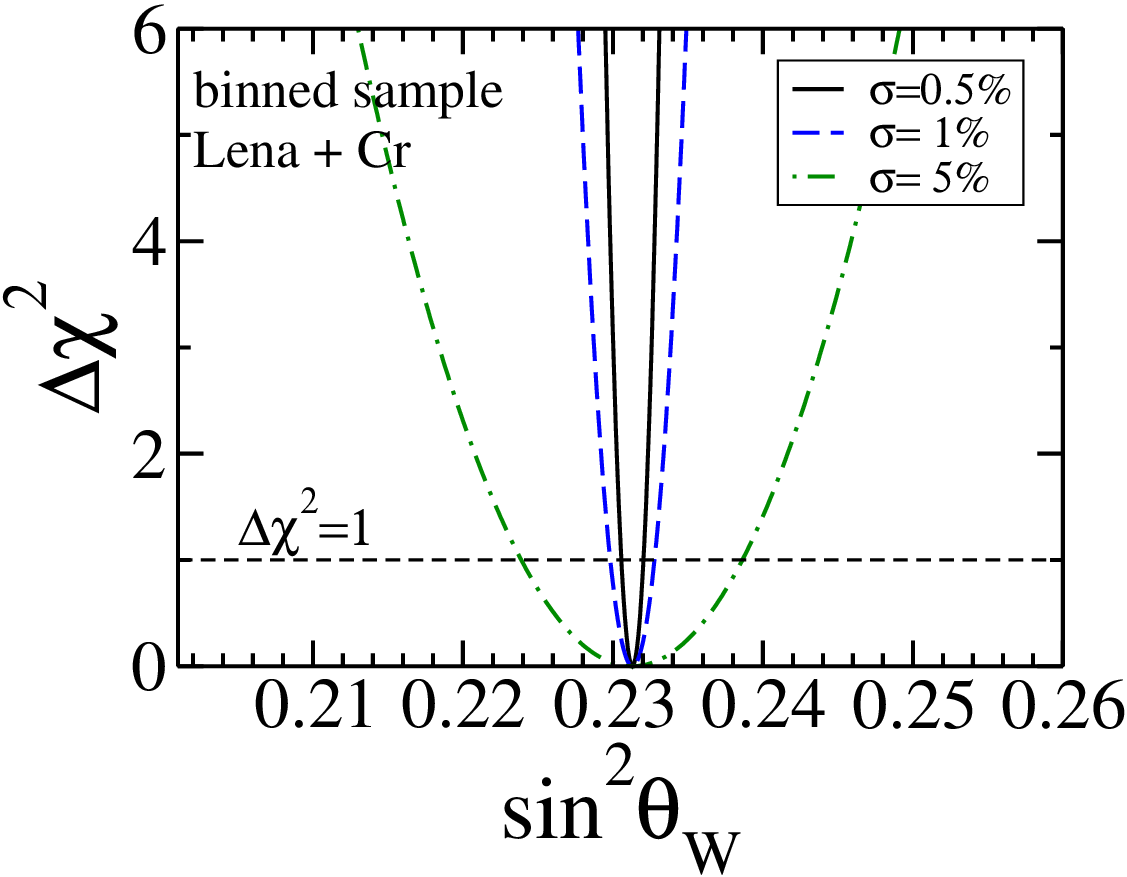}
\includegraphics[width = 0.45 \textwidth]{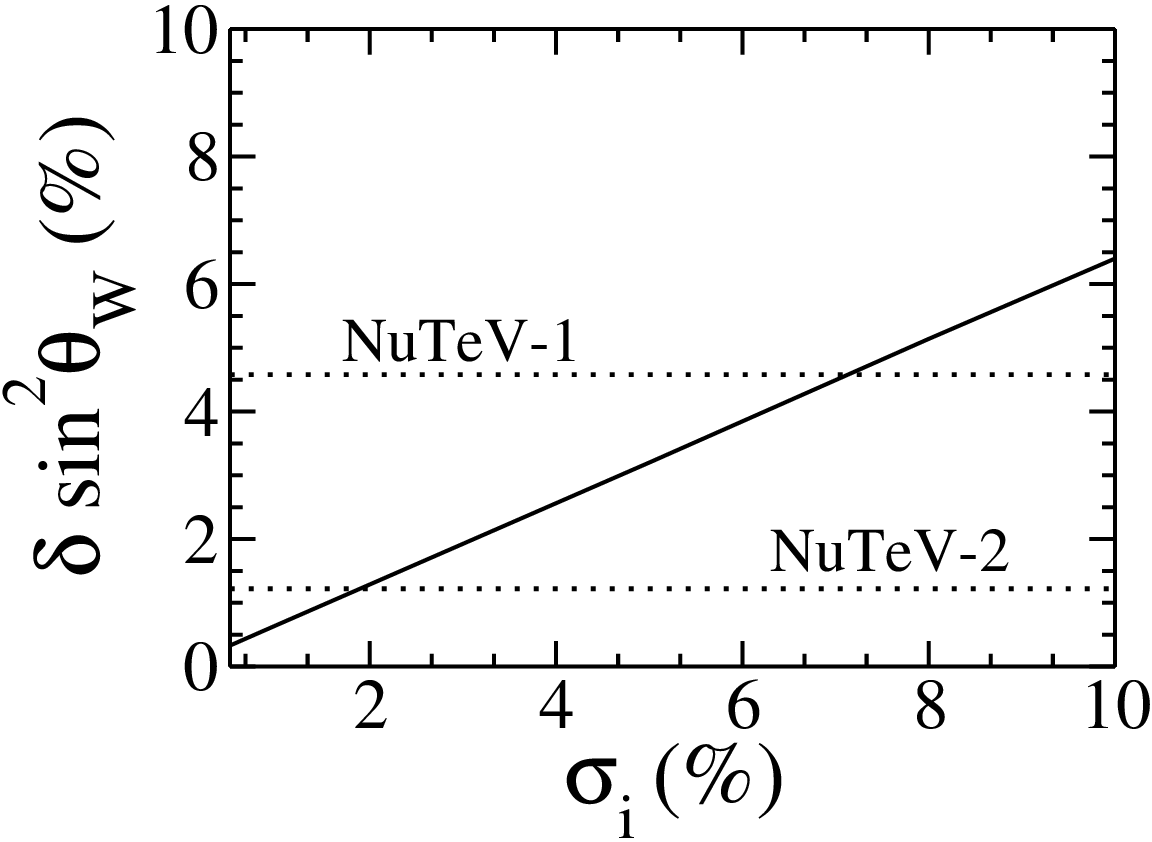}
\caption{Expected sensitivity to the electroweak mixing angle of a
  $^{51}$~Cr neutrino source with the LENA detector. We assume the
  data sample to be divided in seven bins of $50$~keV each and an
  'error' per bin  of $0.5$, $1$ and $5$\%~(left panel). 
  The expected precision (at 68.27\% CL) of the $\sin^2\theta_{\rm
    W}$ determination as a function of the given error in percent is
  given by the tilted line on the right panel. Current NuTeV
  sensitivities are indicated by the two horizontal
  lines~\cite{Bentz:2009yy,Ball:2009mk}.  }
\label{fig:spectrum-Cr}
\end{figure}

A second scenario under consideration is the case of a spectral
binning in the recoil electron energy, $T$. Given the expected energy
resolution in LENA here we imagine that the total sample is split into
seven bins of $50$~keV each.  One performs a $\chi^2$ analysis similar
to the one developed in the previous case and we consider again
different magnitudes for the error per bin (with the same error for
any bin).  We show the results in Fig.~\ref{fig:spectrum-Cr}. 
  One sees how in this case the prospects for a precise determination
  of the electroweak mixing parameter are substantially better than
  those obtained in Fig.~\ref{fig:total-Cr}. Moreover these results
  are less sensitive to potential uncertainties associated to the
  overall normalization of the neutrino flux which could arise, for
  example, in schemes with a light sterile neutrino.
We also compare, as in Fig.~\ref{fig:total-Cr}, the
current NuTeV sensitivities associated to two recent evaluations of
the systematical errors of the
experiment~\cite{Bentz:2009yy,Ball:2009mk}.
One concludes that in this case an experimental error in the LENA
detector of the order of 2\% would suffice to improve the present
sensitivity on the electroweak mixing angle given from the more
precise calculations using the NuTeV measurements.

\subsection{LENA sensitivity to new physics}
\label{sec:sens-new-phys}

Having seen how LENA detector has the potential to improve the
sensitivity on the determination of the electroweak mixing angle we
now turn to the possible search for new physics with LENA. For
definiteness we first consider the sensitivity to the non-standard
neutrino interaction (NSI) parameters that could be generically
associated to the generation of neutrino mass through a low-scale
seesaw
mechanism~\cite{mohapatra:1986bd,gonzalezgarcia:1988rw,valle:1987gv}
or through scalar-boson-mediation~\cite{roulet:1991sm,guzzo:1991hi}.

We assume that a generic effective four-fermion NSI Lagrangian given
as
\begin{equation}
-{\cal L}^{eff}_{\rm NSI} =
\varepsilon_{\alpha \beta}^{fP}{2\sqrt2 G_F} (\bar{\nu}_\alpha \gamma_\rho L \nu_\beta)
( \bar {f} \gamma^\rho P f ) \,,
\label{eq:efflag}
\end{equation}
where $G_F$ is the Fermi constant and $\varepsilon_{\alpha
  \beta}^{fP}$ parametrize the strength of the NSI. This term must be
added to the Standard Model Lagrangian. For laboratory experiments $f$
is a first generation SM fermion ($e, u$ or $d$).  The chiral
projectors $P$ denote $\{L,R=(1\pm\gamma^5)/2\}$, while $\alpha$ and
$\beta$ denote the three neutrino flavors: $e$, $\mu$ and $\tau$.
Our aim is to obtain restrictions on the strength of the NSI
parameters and compare them with those previously reported in the
literature.

In order to illustrate the physics potential of LENA to this type of
scenario we focus on the sensitivity to non universal NSI parameters
for the interaction of neutrinos with electrons.  The differential
cross section for neutrino-electron scattering is therefore modified
due to the presence of the new interactions. In particular, the
coupling constants for the Eq.~(\ref{diff:cross:sec}) will be modified
to have the values
\begin{equation}
g_R \rightarrow g_R+\varepsilon_{e e}^{R};  \,\,\,\,\,\,
g_L \rightarrow g_L+\varepsilon_{e e}^{L}.
\end{equation}
where in the Standard Model couplings $g_{L,R}$ we assume  
  $\sin^2\theta_{\rm W}=0.2313$~\cite{nakamura2010review}.

\begin{figure}
\includegraphics[width = 0.4 \textwidth]{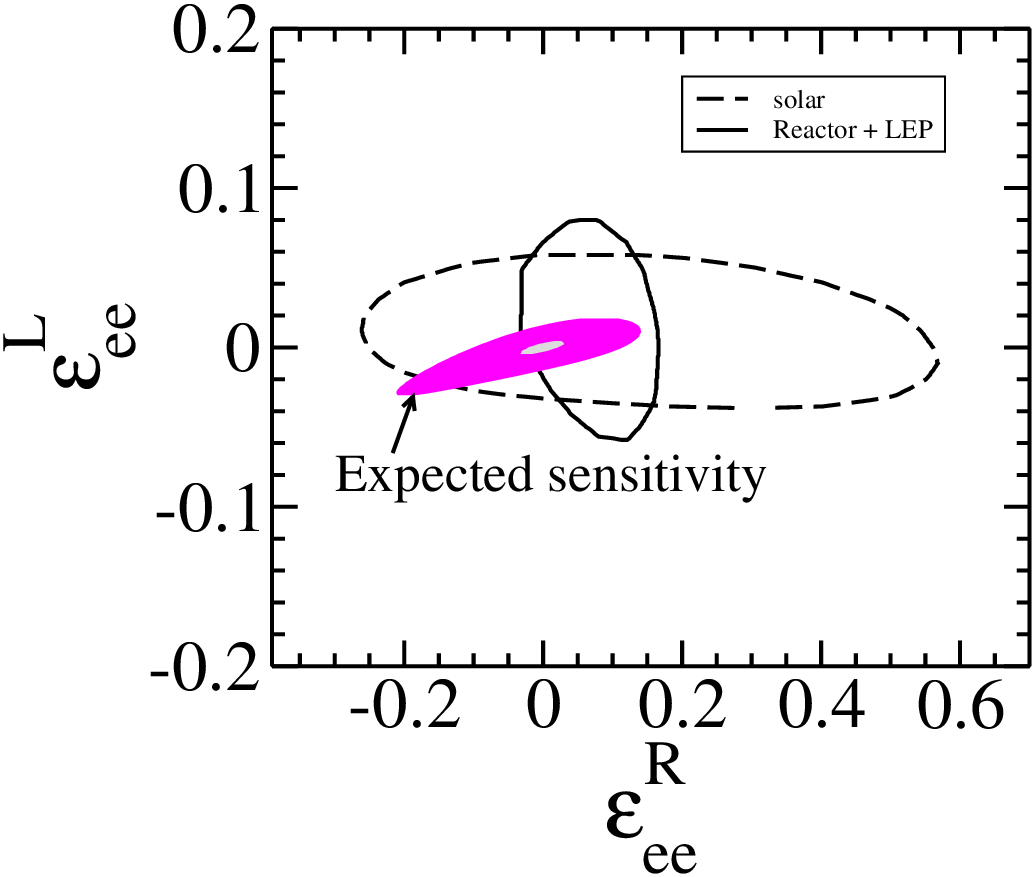}
\includegraphics[width = 0.4 \textwidth]{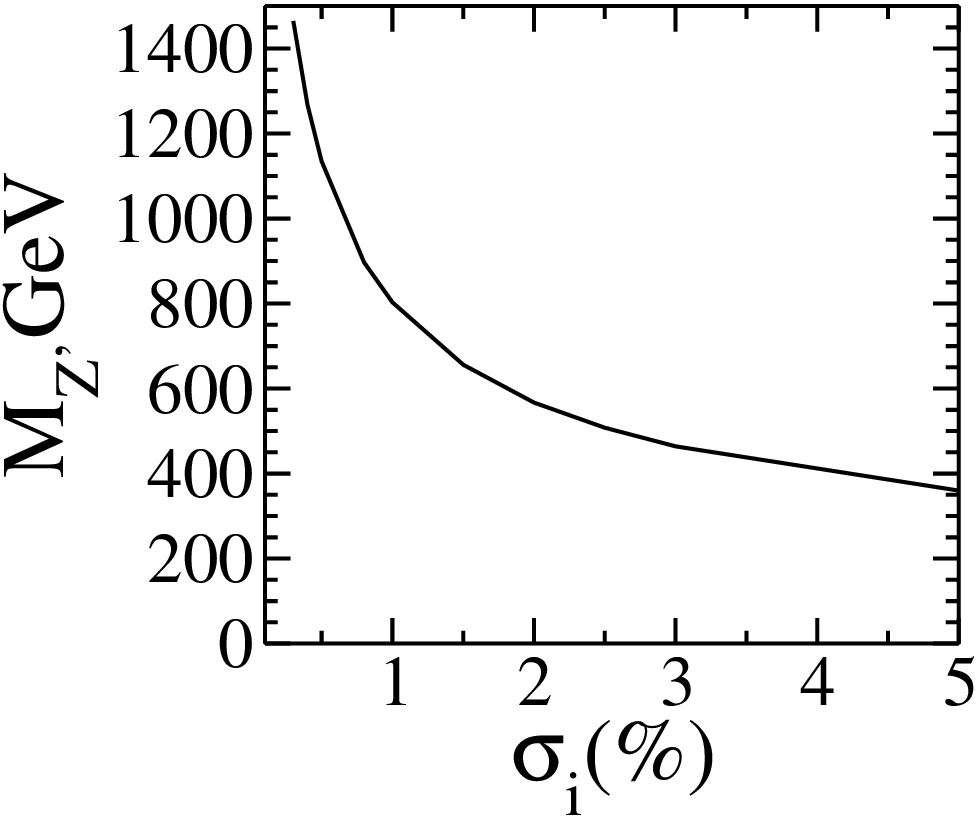}
\caption{Left panel: Expected LENA sensitivity at 90\% CL to
  non-universal NSI using a $^{51}$~Cr neutrino source. The shaded
  areas correspond to a binned data sample divided in seven bins of
  $50$~keV each and an 'error' per bin of either $1$ (grey inner
  region) or $5$\% (magenta outer region). For comparison we show
  current limits to these parameters from an analysis coming from
  solar and KamLAND neutrino data~\cite{Bolanos:2008km} (dashed line)
  as well as from an analysis to the LEP and reactor
  data~\cite{Barranco:2007ej} (solid line).  Right panel: Expected
  sensitivity at 90\% CL to the mass of a new neutral gauge boson
  coupled to lepton number~\cite{valle:1987sq}. In both cases we
  fix the weak mixing angle as $\sin^2\theta_{\rm W}=0.2313$. }
\label{fig:NSI-Cr}
\end{figure}

We have performed a $\chi^2$ analysis, analogous to the one discussed
above, in order to restrict the non universal NSI parameters.  To this
end we study the case of a $^{51}$~Cr source in combination with seven
bins in LENA assuming an error of 0.5\% and 5\% in the measured number
of events per bin.  The results of our analysis are shown in
Fig.~\ref{fig:NSI-Cr} where we show the result for the case of
non-universal NSI~\footnote{Similar results may be obtained also for
  the flavour-changing case.}.
Current constraints, arising either from
solar~\cite{Bolanos:2008km} or LEP+reactor neutrino
experiments~\cite{Barranco:2007ej} are also displayed in the same
figure, for comparison. One can see that even in the most pessimistic
case the LENA sensitivity exceeds the current one. For instance, for a
5\% error in the measured event number per bin the constraint on
$\varepsilon_{e e}^{L}$ would be below a few percent, while for the
case of a $1$\% error the constraint on this parameter will lie below
the percent level.

One can also apply these results to the case of specific theories
beyond the Standard Model involving the presence of an additional
relatively light neutral gauge boson, which may arise in a variety of
scenarios, such as the $E_6$ gauge
group~\cite{mohapatra:1980yp,valle:1987sq,malinsky:2005bi}.  As an
example we take the $\chi$ model discussed in
Ref.~\cite{valle:1987sq}.
The results for the prospected sensitivity of the LENA proposal in
this case are shown in the right panel of Fig.~\ref{fig:NSI-Cr}.
One can see that a constraint in the range from $360$~GeV to $1.1$~TeV
would be attainable depending on the statistics (the assumed error in the
detected event number varying from 0.5\% to 5\% ). It is therefore
clear that for this type of models the sensitivity on the additional
$Z'$ gauge boson mass would be marginal in comparison with the reach
expected at the Large Hadron Collider~\cite{Aad:2011xp}. 
However one may have specific models that predict different couplings
for leptons relative to quarks, suppressing the latter, for example in
leptophilic scenarios~\cite{Fox:2008kb,Ko:2010at}. In this case our
estimated LENA sensitivities would dominate.
Similarly, the LENA proposal would also be useful in restricting
models with trilinear $R$~-parity violating
couplings~\cite{Sessolo:2009ug}.

\section{LENA potential with solar neutrinos} 
\label{sec:lena-potential-with-1}

Here we consider the possibility of performing similar analysis by
using the solar neutrino data collected with the LENA detector, in the
same manner as Borexino. However, in this case the signal would depend
on the neutrino survival probability. Besides, as in
Borexino~\cite{Bellini:2011rx}, the counting per day in the detector
includes the total signal from Beryllium neutrinos as well as from the
background ($C$, $Bi$, $Kr$, etc) which can not be avoided.

Despite these difficulties we have tried to make a forecast of the
sensitivity to the electroweak mixing angle assuming that LENA will be
able to measure the solar neutrino signal in bins of $50$ keV
width. This seems attainable given the expected energy resolution of
the detector. In order to obtain such a forecast we perform an
analysis treating both the survival probability $P_{ee}$ and the
electroweak mixing angle $\sin^2\theta_{\rm W}$ as free parameters. In
order to understand why one may reach a reasonable sensitivity on
$\sin^2\theta_{\rm W}$ despite the dependence on the neutrino survival
probability it is useful to see the expression for the number of
events per bin,
\begin{equation}
N_{i} = n_e\phi_{Be}\Delta t \int^{T_{i+1}}_{T_i}\int  
\left( P_{ee} \frac{d\sigma^{\nu_ee}}{dT} +
(1-P_{ee})\frac{d\sigma^{\nu_{\mu,\tau}e}}{dT} \right)
R(T,T') dT' dT.
\label{diff:cross:sec-osc}
\end{equation}
The differential cross section for the electron-neutrino is given by
Eq.~(\ref{diff:cross:sec}), while for muon or tau neutrinos one has a
similar expression, but with different coupling constants~\footnote{
  In order to make the analytical expressions more transparent we have
  omitted radiative corrections. However they are included in the
  numerical analysis.}
\begin{eqnarray}
\label{gLgR-muon}
g^{\nu_{\mu,\tau}}_L &=& - \frac12 + \sin^2\theta_{\rm W}\\
g^{\nu_{\mu,\tau}}_R &=&    \sin^2\theta_{\rm W} .
\end{eqnarray}
After some simple algebra one sees that the total number of events
can be expressed as
\begin{equation}
N_{i} = n_e\phi_{Be}\Delta t \int^{T_{i+1}}_{T_i}\int  
\left( A(\sin^2\theta_{\rm W}) + B(\sin^2\theta_{\rm W})T+C(\sin^2\theta_{\rm W})T^2 \right)
R(T,T') dT' dT, 
\end{equation}
where the coefficients $A$, $B$ and $C$ are given by 
\begin{eqnarray}
A(\sin^2\theta_{\rm W})&=& 
2(\sin^2\theta_{\rm W})^2-(1-2P_{ee})\sin^2\theta_{\rm W}+\frac{1}{4}, \\  
\nonumber
B(\sin^2\theta_{\rm W})&=&
\frac{ \sin^2\theta_{\rm W} m_e}{E^2} (\frac{1}{2}-\sin^2\theta_{\rm W}-P_{ee})
-\frac{2(\sin^2\theta_{\rm W})^2}{E},\\  
\nonumber
C(\sin^2\theta_{\rm W})&=& 
\frac{(\sin^2\theta_{\rm W})^2}{E}. \\  \nonumber
\end{eqnarray}
One sees that the effect of the neutrino survival probability in the
shape of the spectrum is minimal for the case of $P_{ee}\simeq 0.5$
which is close to the expected value in this region. Therefore, the
main effect of $P_{ee}$ will be to reduce the total number of events
while the effect in the shape of the spectrum will be mild.  For
instance, the coefficient $C$ of the quadratic term in $T$ has no
dependence on $P_{ee}$.

Notice also that for the $^7$Be line we have in principle a fixed
neutrino energy, $E_\nu = 0.862$~MeV, and therefore the survival
probability is computed also for this energy value and hence there is
no need to convolute the signal over an energy range. These features
make the analysis more transparent.

\begin{figure}
\includegraphics[width = 0.45 \textwidth]{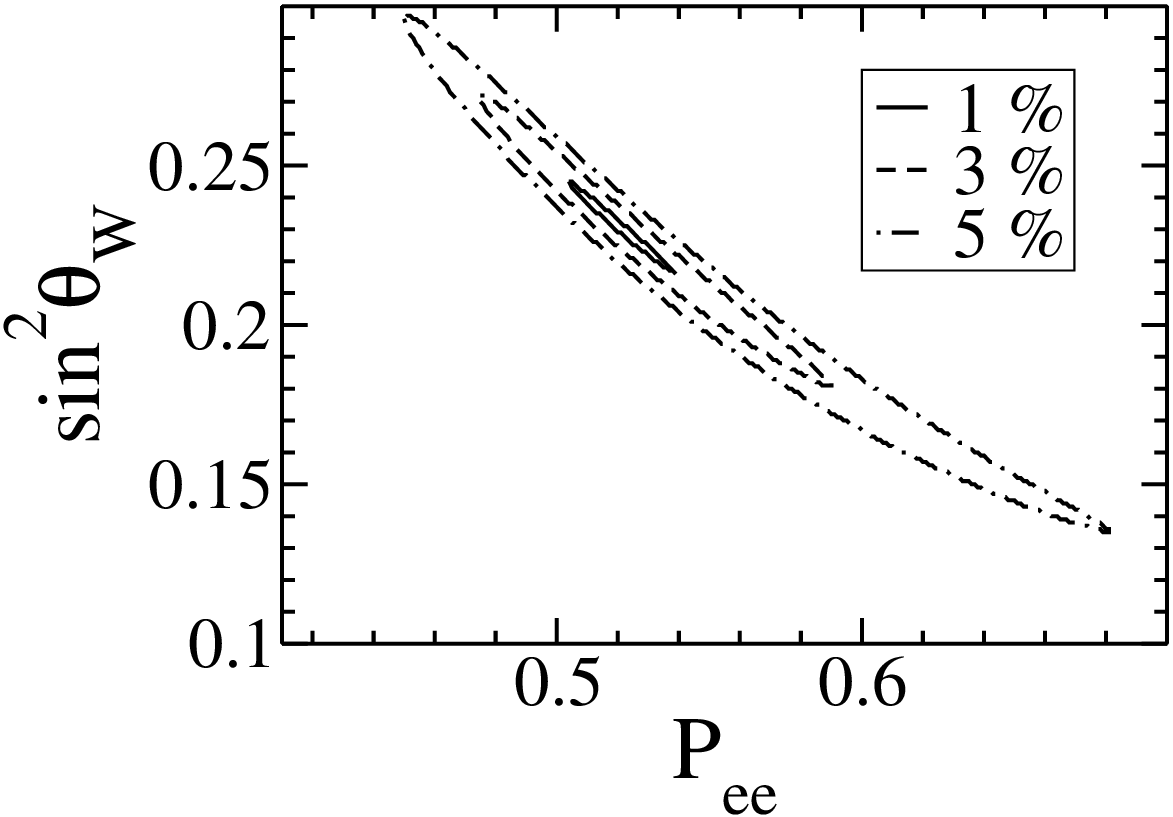}
\caption{Expected sensitivity (at 68.27\% CL) of the LENA detector to
  the solar neutrino Beryllium signal, assuming a $50$~keV recoil
  energy binning and an error per bin of $1$, $3$ and $5$\%.}
\label{fig:spectrum-Be}
\end{figure}

Although it is difficult to forecast how well the LENA collaboration
will measure the Beryllium line, we would like to give at least an
estimate of the LENA sensitivity to the electroweak mixing
parameter.
To this end we proceed with a $\chi^2$ analysis similar to the one
introduced in the previous section, but given this time in terms of
two parameters, the electroweak mixing angle $\sin^2\theta_{\rm W}$
and the neutrino survival probability $P_{ee}$.
In Fig.~\ref{fig:spectrum-Be} we show the expected sensitivity to
the neutrino survival probability and the electroweak mixing angle
corresponding to a $50$~keV recoil energy binning and an error of
$1,3$ and $5$\% for each bin.
One can see that for a relatively precise determination of the
Beryllium spectrum there will be a reasonable sensitivity to
$\sin^2\theta_{\rm W}$. For instance, in the optimistic case of an
error in the event number per bin of $1$\%, and despite the
correlation with the survival probability, there would be a
sensitivity to the electroweak mixing parameter of the order $\sim
6$\%.

\section{Conclusions}

We have studied the potential of the LENA proposal for electroweak
measurements in combination with a radioactive Chromium source.
We showed how it could indeed provide a precise measurement of the
electroweak mixing angle in a region of energy that is not easy to
study with other experiments. We have also discussed some possible
applications of LENA to probe physics beyond the Standard Model, such
as non-standard neutrino interactions, and the possible existence of 
new electroweak neutral gauge bosons.

We also discussed the potential of the LENA detector for the solar
Beryllium signal. Although in this case the goal will be more
challenging, our results indicate that it would be worthwhile to
perform a more realistic simulation of the LENA detector in order to
determine more accurately its potential, by taking advantage of
current Borexino spectral results. Although current data may be poor
for this type of physics, they may be helpful to obtain better
estimates of the future sensitivities attainable in LENA.
Other proposals to probe the electroweak mixing
parameter~\cite{Agarwalla:2010ty,Balantekin:2005md,Conrad:2004gw} have
also been considered. 
The accuracy in the determination of the electroweak mixing parameter
expected in LENA lies below the percent level for the most optimistic
expectations on the systematical error.
These results indicate that LENA holds good prospects, quite
competitive with the alternatives.

\vfill

\section{Acknowledgments}

This work was supported by the Spanish MICINN under grants
FPA2008-00319/FPA, FPA2011-22975 and MULTIDARK CSD2009-00064
(Consolider-Ingenio 2010 Programme), by Prometeo/2009/091 (Generalitat
Valenciana), by the EU ITN UNILHC PITN-GA-2009-237920 and by CONACyT grant 
166639 (Mexico). M. A. T. acknowledges financial support from CSIC under the
JAE-Doc programme. O. G. M. is also supported by EPLANET.

  \renewcommand{\baselinestretch}{1.2}
\bibliographystyle{h-physrev4}

\end{document}